\documentclass[12pt]{article}

\usepackage{amsmath,amssymb}
\usepackage{graphicx}
\usepackage{caption}
\usepackage{color}
\usepackage{dcolumn}
\usepackage{bm}
\usepackage[numbers,super,comma,sort&compress]{natbib}
\usepackage{float}
\usepackage{multirow}

\usepackage{threeparttable}
\usepackage{booktabs}
\usepackage{listings}
\usepackage{ulem}
\usepackage{authblk}
\usepackage{cprotect}

\allowdisplaybreaks     
\graphicspath{{image/}}

\definecolor{deepred}{rgb}{0.5,0,0}
\definecolor{deepgreen}{rgb}{0,0.4,0}
\definecolor{deepblue}{rgb}{0,0,0.6}

\lstnewenvironment{pyscf}[1][]
{\lstset{language=Python,
escapeinside={(*@}{@*)},
basicstyle=\linespread{1.25}\footnotesize\ttfamily,        
commentstyle=\ttfamily\color{deepblue},
frame=tb, 
keepspaces=true,                 
columns=fullflexible,
keywordstyle=\color{deepred},
stringstyle=\color{deepgreen},
showspaces=false,
showstringspaces=false,
showtabs=false,
tabsize=2,
moredelim=[is][\underbar]{**}{**},
}}{}

\makeatletter

\newcommand{\Runum}[1]{\expandafter\@slowromancap\romannumeral #1@}
\makeatother

\title{PySCF: The Python-based Simulations of Chemistry Framework}

\begin{document}

\author[1]{Qiming Sun\thanks{osirpt.sun@gmail.com}}
\author[2]{Timothy C. Berkelbach}
\author[3,4]{Nick S. Blunt}
\author[5]{George H. Booth}
\author[1,6]{Sheng Guo}
\author[1]{Zhendong Li}
\author[7]{Junzi Liu}
\author[1,6]{James D. McClain}
\author[1,6]{Elvira R. Sayfutyarova}
\author[8]{Sandeep Sharma}
\author[9]{Sebastian Wouters}
\author[1]{Garnet~Kin-Lic~Chan\thanks{gkc1000@gmail.com}}
\affil[1]{Division of Chemistry and Chemical Engineering, California Institute of Technology, Pasadena CA 91125, USA}
\affil[2]{Department of Chemistry and James Franck Institute, University of Chicago, Chicago, Illinois 60637, USA}
\affil[3]{Chemical Science Division, Lawrence Berkeley National Laboratory, Berkeley, California 94720, USA}
\affil[4]{Department of Chemistry, University of California, Berkeley, California 94720, USA}
\affil[5]{Department of Physics, King's College London, Strand, London WC2R 2LS, United Kingdom}
\affil[6]{Department of Chemistry, Princeton University, Princeton, New Jersey 08544, USA}
\affil[7]{Institute of Chemistry Chinese Academy of Sciences, Beijing 100190, P. R. China}
\affil[8]{Department of Chemistry and Biochemistry, University of Colorado Boulder, Boulder, CO 80302, USA}
\affil[9]{Brantsandpatents, Pauline van Pottelsberghelaan 24, 9051 Sint-Denijs-Westrem, Belgium}

\date{}

\maketitle

\begin{center}

\subsubsection*{Abstract}


\textsc{PySCF} is a general-purpose electronic structure platform designed from the ground
up to emphasize code simplicity, so as to facilitate new method development and enable
flexible computational workflows.  The package provides a wide range of tools to
support simulations of finite-size systems, extended systems with periodic boundary
conditions, low-dimensional periodic systems, and custom Hamiltonians, using mean-field
and post-mean-field methods with standard Gaussian basis functions.  To ensure ease of
extensibility, \textsc{PySCF} uses the Python language to implement almost all of its
features, while computationally critical paths are implemented with heavily optimized C
routines.  Using this combined Python/C implementation, the package is as efficient as the
best existing C or Fortran based quantum chemistry programs.  In this paper we document
the capabilities and design philosophy of the current version of the \textsc{PySCF}
package.
\end{center}

\clearpage

\makeatletter
\renewcommand\@biblabel[1]{#1.}
\makeatother
\bibliographystyle{apsrev}

\renewcommand{\baselinestretch}{1.5}
\normalsize

\clearpage

\section{\sffamily \Large INTRODUCTION}

The Python programming language is playing an increasingly important role in scientific
computing.  As a high level language, Python supports rapid development practices and easy
program maintenance.  While programming productivity is hard to measure, it is commonly
thought that it is more efficient to prototype new ideas in Python, rather than in
traditional low-level compiled languages such as Fortran or C/C++. 
Further, through the use of the many high-quality numerical libraries available in Python
-- such as \textsc{NumPy}\cite{numpy}, \textsc{SciPy}\cite{scipy}, and \textsc{MPI4Py}\cite{mpi4py}
-- Python programs can perform at competitive levels with optimized Fortran and C/C++
programs, including on large-scale computing architectures.

There have been several efforts in the past to incorporate Python into electronic structure
programs.
Python has been widely adopted in a scripting role: 
the \textsc{Psi4}\cite{psi4} quantum chemistry package uses a custom ``Psithon'' dialect to
drive the underlying C++ implementation, while general simulation environments
such as \textsc{ASE}\cite{ase} and \textsc{PyMatGen}\cite{Shyue2013} provide Python frontends to multiple
quantum chemistry and electronic structure packages, to organize complex workflows\cite{pyadf}. Python
has also proved popular for implementing symbolic second-quantized algebra and code
generation tools, such as the Tensor Contraction Engine\cite{Hirata2003} and the
SecondQuantizationAlgebra library\cite{Neuscamman2009,Saitow2013}.

In the above cases, Python has been used as a supporting language, with the underlying
quantum chemistry algorithms implemented in a compiled language.  However, Python has also
seen some use as a primary implementation language for electronic structure methods.
\textsc{PyQuante}\cite{pyquante} was an early attempt to implement a Gaussian-based
quantum chemistry code in Python, although it did not achieve speed or functionality
competitive with typical packages.  Another early effort was the \textsc{GPAW}\cite{gpaw}
code, which implements the projector augmented wave formalism for density functional
theory, and which is still under active development in multiple groups.  Nonetheless, it
is probably fair to say that using Python as an implementation language, rather than a
supporting language, remains the exception rather than the rule in modern quantum
chemistry and electronic structure software efforts.



With the aim of developing a new highly functional, high-performance computing toolbox
for the quantum chemistry of molecules and materials implemented primarily in the
Python language, we started the open-source project ``Python-based Simulations of Chemistry Framework'' (\textsc{PySCF}) in 2014.
The program was initially ported from our quantum chemistry density matrix embedding 
theory (DMET) project\cite{Sun2014} and contained only the Gaussian integral interface, a
basic Hartree-Fock solver, and a few post-Hartree-Fock components required by DMET.
In the next 18 months, multi-configurational self-consistent-field (MCSCF), density
functional theory and coupled cluster theory, as well as relevant modules for
molecular properties, were added into the package.
In 2015, we released the first stable version, PySCF~1.0, wherein we codified our primary
goals for further code development: to produce a package that emphasizes simplicity,
generality, and efficiency, in that order.
As a result of this choice, most functions in PySCF are written purely in Python, with a very 
limited amount of C code only for the most time-critical parts.
The various features and APIs are designed and implemented in the simplest and most
straightforward manner, so that users can easily modify the source code to meet their own
scientific needs and workflow.
However, although we have favored algorithm accessibility and extensibility over
performance, we have found that the efficient use of numerical Python libraries allows
\textsc{PySCF} to perform at least as fast as the best existing quantum chemistry
implementations. 
In this article, we highlight the current capabilities and design philosophy of the
\textsc{PySCF} package.

\section{\sffamily \Large CAPABILITIES}

Molecular electronic structure methods are typically the main focus of quantum chemistry
packages. We have put significant effort towards the production of a stable, feature-rich
and efficient molecular simulation environment in \textsc{PySCF}.
In addition to molecular quantum chemistry methods, \textsc{PySCF} also provides a wide
range of quantum chemistry methods for extended systems with periodic boundary conditions
(PBC).  Table \ref{tab:features} lists the main electronic structure
methods available in the \textsc{PySCF} package.
More detailed descriptions are presented in Section \ref{sec:scf} - Section \ref{sec:pbc}.
Although not listed in the table, many auxiliary tools for method development are also part of the package.
They are briefly documented in Section \ref{sec:aoints} - Section \ref{sec:tools}.

\subsection{\sffamily \normalsize Self-consistent field methods}
\label{sec:scf}
Self-consistent field (SCF) methods are the starting point for most electronic structure
calculations. In \textsc{PySCF}, the SCF module includes implementations of Hartree-Fock
(HF) and density functional theory (DFT) for restricted, unrestricted, closed-shell and
open-shell Slater determinant references.  A wide range of predefined exchange-correlation (XC) functionals
are supported through a general interface to the \textsc{Libxc}\cite{libxc} and
\textsc{Xcfun}\cite{xcfun} functional libraries.  Using the interface, as shown
in Figure \ref{fig:xc}, one can easily customize the XC functionals in DFT calculations.
\textsc{PySCF} uses the \textsc{Libcint}\cite{Sun2015} Gaussian
integral library, written by one of us (QS) as its integral engine.
In its current implementation, the SCF program can handle over 5000 basis functions on a
single symmetric multiprocessing (SMP) node without any approximations to the integrals.
To obtain rapid convergence in the SCF iterations, we have also developed a second order
co-iterative augmented Hessian (CIAH) algorithm for orbital optimization~\cite{Sun2016a}.
Using the direct SCF technique with the CIAH algorithm, we are able to converge
a Hartree-Fock calculation for the open-shell molecule Fe(II)-porphine (2997
AOs) on a 16-core node in one day.

\subsection{\sffamily \normalsize Post SCF methods}
\label{sec:posthf}

Single-reference correlation methods can be used on top of the HF or DFT references,
including M\o ller-Plesset second-order perturbation theory (MP2), configuration
interaction, and coupled cluster theory. 

Canonical single-reference coupled cluster theory has been implemented with 
single and double excitations (CCSD)\cite{Koch1996} and with perturbative triples
[CCSD(T)].
The associated derivative routines include CCSD and CCSD(T) density matrices, CCSD and
CCSD(T) analytic gradients, and equation-of-motion CCSD for the ionization potentials,
electron affinities, and excitation energies
(EOM-IP/EA/EE-CCSD)\cite{Sekino1984,Nooijen1995,Bartlett2003}.
The package contains two complementary implementations of each of these methods.
The first set are straightforward spin-orbital and spatial-orbital implementations, which
are optimized for readability and written in pure Python using syntax of the 
\textsc{NumPy} \verb$einsum$ function (which can use either the default \textsc{Numpy} implementation
or a custom \texttt{gemm}-based version) for tensor contraction.
These implementations are easy for the user to modify.  A second
spatial-orbital implementation has been intensively optimized to minimize
dataflow and uses asynchronous I/O and a threaded \texttt{gemm} function for efficient
tensor contractions.  For a system of 25 occupied orbitals and 1500 virtual orbitals 
(H$_{50}$ with cc-pVQZ basis), the latter CCSD implementation takes less than 3 hours to
finish one iteration using 28 CPU cores. 

The configuration interaction code implements two solvers: a solver for configuration
interaction with single and double excitations (CISD), and a determinant-based full
configuration interaction (FCI) solver\cite{Handy1984} for fermion, boson or coupled
fermion-boson Hamiltonians.  The CISD solver has a similar program layout
to the CCSD solver.  The FCI solver additionally implements the spin-squared operator,
second quantized creation and annihilation operators (from which arbitrary second
quantized algebra can be implemented), functions to evaluate the density matrices and
transition density matrices (up to fourth order), as well as a function to evaluate the
overlap of two FCI wavefunctions in different orbital bases.  The FCI solver is
intensively optimized for multi-threaded performance.  It can perform one matrix-vector
operation for 16 electrons and 16 orbitals using 16 CPU cores in 30 seconds.

\subsection{\sffamily \large Multireference methods}

For multireference problems, the \textsc{PySCF} package provides the complete active space self
consistent field (CASSCF) method\cite{Jensen1987,Werner1985} and N-electron
valence perturbation theory (NEVPT2)\cite{Malrieu2001,Guo2016}.
When the size of the active space exceeds the capabilities of the
conventional FCI solver, one can switch to  external variational solvers
such as a density matrix renormalization group (DMRG) program\cite{Sharma2012,CheMPS2cite1,casscf2016}
or a full configuration interaction quantum Monte Carlo (FCIQMC) program\cite{Booth2009,Thomas2015}.
Incorporating  external solvers into the CASSCF optimizer widens
the range of possible applications, while raising new challenges for an efficient CASSCF
algorithm.
One challenge is the communication between the external solver and the orbital
optimization driver; communication must be limited to quantities
that are easy to obtain from the external solver.
A second challenge is the cost of handling quantities associated with the active space;
for example, as the active space becomes large, the memory required
to hold integrals involving active labels can easily exceed available memory.
Finally, any approximations introduced in the context of the above two challenges
should not interfere with the quality of convergence of the CASSCF optimizer.

To address these challenges,
we have implemented a general AO-driven CASSCF optimizer\cite{casscf2016} that
provides second order convergence and which may easily be combined with a wide
variety of external variational solvers, including DMRG, FCIQMC and their
state-averaged solvers.
Only the 2-particle density matrix and Hamiltonian integrals
are communicated between the CASSCF driver and the external CI solver.
Further, the AO-driven algorithm has a low memory and I/O footprint. The current
implementation supports
calculations with 3000 basis functions and 30--50 active orbitals
on a single SMP node with 128 GB memory, without any approximations to the AO integrals.

A simple interface is provided to use an external solver in 
multiconfigurational calculations.
Figure \ref{fig:dmrgscf} shows how to perform a
DMRG-CASSCF calculation by replacing the \verb$fcisolver$ attribute of the CASSCF
method.
DMRG-SC-NEVPT2\cite{Guo2016}, and ic-MPS-PT2 and
ic-MPS-LCC\cite{Sharma2016} methods are also available through the interface to the DMRG
program package \textsc{Block}\cite{DMRG2002,DMRG2004,DMRG2008,Sharma2012}, and the \textsc{ic-MPS-LCC} program of
Sharma~\cite{Sharma2016}.

\subsection{\sffamily \normalsize Molecular properties}

At the present stage, the program can compute molecular properties 
such as analytic nuclear gradients, analytic nuclear Hessians, and NMR
shielding parameters at the SCF level.  The CCSD and CCSD(T) modules include solvers for
the $\Lambda$-equations.  As a result, we also provide one-particle and two-particle density
matrices, as well as the analytic nuclear gradients, for the CCSD and CCSD(T)
methods\cite{Bartlett1989}.

For excited states, 
time-dependent HF (TDHF) and time-dependent DFT (TDDFT)
are implemented on top of the SCF module.  The relevant analytic nuclear
gradients are also programmed\cite{Furche2002}.
The CCSD module offers another option to obtain
excited states using the EOM-IP/EA/EE-CCSD methods.  The third option
to obtain excited states
is through the multi-root CASCI/CASSCF solvers, optionally followed by the MRPT tool chain.  
Starting from the multi-root CASCI/CASSCF solutions, the program can compute
the density matrices of all the states and the transition density matrices
between any two states.  One can contract these density matrices with
specific AO integrals to obtain different first order molecular properties.

\subsection{\sffamily \normalsize Relativistic effects}

Many different relativistic treatments are available in \textsc{PySCF}.
Scalar relativistic effects can be added to all SCF and post-SCF methods through relativistic
effective core potentials (ECP)\cite{ecp2006} or the all-electron spin-free X2C\cite{Liu2009}
relativistic correction.
For a more advanced treatment, \textsc{PySCF} also provides 4-component relativistic
Hartree-Fock and no-pair MP2 methods with Dirac-Coulomb, Dirac-Coulomb-Gaunt, and
Dirac-Coulomb-Breit Hamiltonians.
Although not programmed as a standalone module, the no-pair
CCSD electron correlation energy can also be computed with the straightforward
spin-orbital version of the CCSD program.
Using the 4-component Hamiltonian, molecular properties including analytic
nuclear gradients and NMR shielding parameters are available at the mean-field level\cite{Cheng2009}.

\subsection{\sffamily \normalsize Orbital localizer and result analysis}

Two classes of orbital localization methods are available in the package.
The first emphasizes  the atomic character of the basis functions.
The relevant localization functions can generate intrinsic atomic orbitals
(IAO)\cite{Knizia2013}, natural atomic orbitals (NAO)\cite{Weinhold1988},
and meta-L\"owdin orbitals\cite{Sun2014} based on orbital projection and
orthogonalization.
With these AO-based local orbitals, charge distributions can be properly
assigned to atoms in  population analysis\cite{Knizia2013}.
In the \textsc{PySCF} population analysis code, meta-L\"owdin orbitals are the
default choice.

The second class, represented by Boys-Foster, Edmiston-Ruedenberg, and
Pipek-Mezey localization, require minimizing (or maximizing) the dipole, the
Coulomb self-energy, or the atomic charges, to obtain the optimal localized orbitals.
The localization routines can take arbitrary orthogonal input orbitals and call
the CIAH algorithm to rapidly converge the solution.
For example, using 16 CPU cores, it takes 3 minutes to localize 1620 HF unoccupied
orbitals for the C$_{60}$ molecule using Boys localization.

\begin{figure}[htp]
\centering
\begin{minipage}[c]{.6\textwidth}
\centering
\begin{pyscf}
from pyscf import gto, scf, lo, tools
mol = gto.Mole(atom=open('c60.xyz').read(),
               basis='ccpvtz')
mf = scf.RHF(mol).run()
orb = lo.Boys(mol).kernel(mf.mo_coeff[:,:180])
tools.molden.from_mo(mol, 'c60.molden', orb)

# Invoke Jmol to plot the orbitals
with open('c60.spt', 'w') as f:
  f.write('load c60.molden; isoSurface MO 002;\n')
import os
os.system('jmol c60.spt')
\end{pyscf}
\end{minipage}
\begin{minipage}[c]{.3\textwidth}
  \hspace{-4ex}
  \includegraphics[width=1.\textwidth]{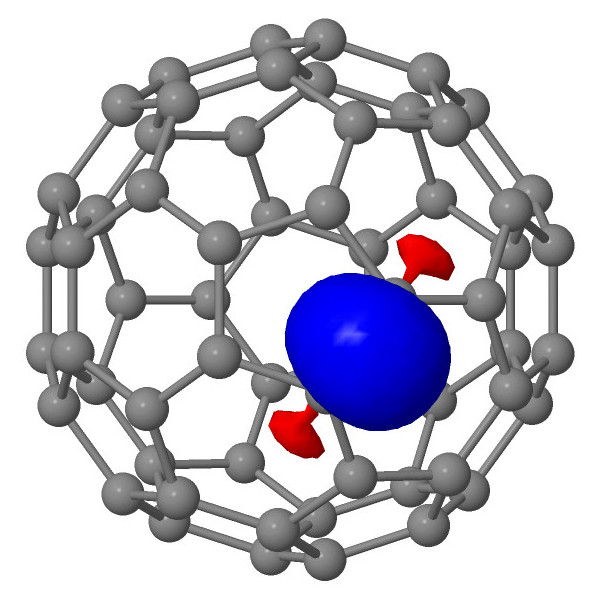}
\end{minipage}
  \caption{Example to generate localized orbitals and to plot them in Jmol.}
  \label{fig:localorb}
\end{figure}

A common task when analysing the results of an electronic structure calculation
is to visualize the orbitals.
Although \textsc{PySCF} does not have a visualization tool itself, it provides a module
to convert the given orbital coefficients to the \verb$molden$\cite{molden} format which can
be read and visualized by other software, e.g. \textsc{Jmol}\cite{jmol}.
Figure \ref{fig:localorb} is an example to run Boys localization for the
C$_{60}$ HF occupied orbitals and to generate the orbital surfaces of the localized
$\sigma$-bond orbital in a single Python script.

\subsection{\sffamily \normalsize Extended systems with periodic boundary conditions}
\label{sec:pbc}
PBC implementations typically use either plane
waves\cite{abinit,nwchem,quantumespresso,wien2k} or local atomic
functions\cite{crystal14,cryscor,gpaw,blum2009ab,artacho2008siesta,cp2k} as the underlying orbital basis.
The PBC implementation in \textsc{PySCF} uses the local basis formulation, specifically
crystalline orbital Gaussian basis functions $\phi$, expanded in terms of a lattice sum
over local Gaussians $\chi$
\begin{equation*}
  \phi_{\bm{k},\chi}(\bm{r}) = \sum_{\bm{T}}
        e^{i\bm{k}\cdot\bm{T}} \chi(\bm{r} - \bm{T})
\end{equation*}
where $\bm{k}$ is a vector in the first Brillouin zone and $\bm{T}$ is a lattice
translational vector.
We use a pure Gaussian basis in our PBC implementation for two reasons: to 
simplify the development of post-mean-field methods for extended systems and to
have a seamless interface and direct comparability to
finite-sized quantum chemistry calculations.  Local bases are favourable for
post-mean-field methods because they are generally quite compact, resulting in small
virtual spaces~\cite{booth2016plane}, and further allow locality to be exploited.  
Due to the use of local bases, various boundary conditions can be easily
applied in the PBC module, from  zero-dimensional systems (molecules) to
extended one-, two- and three-dimensional periodic systems.

The PBC module supports both all-electron and pseudopotential calculations.  Both
separable pseudopotentials (e.g. Goedecker-Teter-Hutter (GTH)
pseudopotentials~\cite{GTH,HGH}) and non-separable pseudopotentials (quantum chemistry ECPs and
Burkatzi-Filippi-Dolg pseudopotentials\cite{BFD2007}) can be
used.  
In the separable pseudopotential implementation, the associated orbitals and densities are
guaranteed to be smooth, allowing a grid-based treatment that uses discrete fast Fourier
transforms~\cite{cp2k,McClain2017}.
In both the pseudopotential and all-electron PBC calculations, 
Coulomb-based integrals are handled via density fitting as described in Section \ref{sec:df}.


The PBC implementation is organized in direct correspondence to the molecular implementation.
We implemented the same function interfaces as in the molecular code,
with analogous module and function names.
Consequently, methods defined in the molecular part of the code can be seamlessly mixed 
with the PBC functions without modification, especially in $\Gamma$-point calculations 
where the PBC wave functions are real.
Thus, starting from PBC $\Gamma$-point mean-field orbitals, one can,
for example, carry out CCSD, CASSCF, TDDFT, etc.~calculations using
the molecular implementations.
We also introduce specializations of the PBC methods to support
$k$-point (Brillouin zone) sampling. 
The $k$-point methods slightly
modify the $\Gamma$-point data structures, but inherit from and reuse
almost all of the $\Gamma$-point functionality.
Explicit $k$-point sampling is supported at the HF and DFT level, and on top of this we
have also implemented $k$-point MP2, CCSD, CCSD(T) and EOM-CCSD methods\cite{McClain2017}, with
optimizations to carefully distribute work and data across cores.
On 100 computational cores, mean-field
simulations including unit cells with over 100 atoms, or $k$-point CCSD calculations with
over 3000 orbitals, can be executed without difficulty.

\subsection{\sffamily \normalsize General AO integral evaluator and J/K builds}
\label{sec:aoints}
Integral evaluation forms the foundation of Gaussian-based electronic
structure simulation.
The general integral evaluator library \textsc{Libcint} supports a wide range of
GTO integrals, and \textsc{PySCF} exposes simple APIs to access the \textsc{Libcint} integral
functions.
As the examples in Figure~\ref{fig:ao:ints} show, 
the \textsc{PySCF} integral API allows users to access AO integrals either in a
giant array or in individual shells with a single line of Python code. The
integrals provided include,
\begin{itemize}
  \item integrals in the basis of Cartesian, real-spherical and $j$-adapted spinor GTOs;
  \item arbitrary integral expressions built from $\bm{r}$, $\bm{p}$, and
    $\sigma$ polynomials;
  \item 2-center, 3-center and 4-center 2-electron integrals for the Coulomb
    operator $1/r_{12}$,  range-separated Coulomb operator
    $\mathrm{erf}(\omega r_{12})/r_{12}$,
    Gaunt interaction, and Breit interaction.
\end{itemize}

Using the general AO integral evaluator, the package provides a general
AO-driven J/K contraction function.
J/K-matrix construction involves a contraction over a high order tensor
(e.g. 4-index 2-electron integrals $(ij|kl)$) and a low order tensor (e.g. the 2-index
density matrix $\gamma$)
\begin{gather*}
  J_{ij} = \sum_{kl} (ij|kl) \gamma_{kl} \\
  K_{il} = \sum_{jk} (ij|kl) \gamma_{jk}
\end{gather*}
When both tensors can be held in memory, the \textsc{Numpy} package offers a convenient
tensor contraction function \verb$einsum$ 
to quickly construct J/K matrices. 
However, it is common for the high order tensor to be too large to fit into the
available memory.
Using the Einstein summation notation of the \textsc{Numpy} \verb$einsum$ function,
our AO-driven J/K contraction implementation offers the capability to contract the
high order tensor (e.g.~2-electron integrals or their high order derivatives) with
multiple density matrices, with a small memory footprint.
The J/K contraction function also supports subsystem contraction, in
which the 4 indices of the 2-electron integrals are distributed over different segments
of the system which may or may not overlap with each other.
This subsystem contraction is particularly useful in two scenarios:
in fragment-based methods, where the evaluation of Coulomb or exchange energies
involves integral contraction over different fragments, and
in parallel algorithms, where one partitions the J/K contraction into
small segments and distributes them to different computing nodes.


\subsection{\sffamily \normalsize General integral transformations}

Integral transformations are another fundamental operation found in quantum
chemistry programs. A common kind of integral transformation is to transform the 4 indices of the
2-electron integrals by 4 sets of different orbitals.
To satisfy this need, we designed a general integral transformation function to
handle the arbitrary AO integrals provided by the \textsc{Libcint} library and arbitrary kinds of orbitals.
To reduce  disk usage, we use permutation symmetry over $i$ and $j$, $k$ and $l$ in
$(ij|kl)$ whenever  possible for real integrals.

Integral transformations involve high computational and I/O costs.  
A standard approach to reduce these costs involves precomputation to reduce integral costs and data
compression to increase I/O throughput.  However, we have not adopted such an
optimization strategy in our implementation because it is against the objective of simplicity
for the \textsc{PySCF} package.  In our implementation, initialization is not 
required for the general integral transformation function.  Similarly to the AO integral API, the
integral transformation can thus be launched with one line of Python code.  In the
integral data structure, we store the transformed integrals by chunks in the HDF5
format without compression.  This choice has two advantages.  First, it allows
for fast indexing and hyperslab selection for subblocks of the integral array.
Second, the integral data can be easily accessed by other program packages
without any overhead for parsing the integral storage protocol.

\subsection{\sffamily \normalsize Density fitting}
\label{sec:df}
The density fitting (DF) technique is implemented for both  finite-sized
systems and crystalline systems with periodic boundary conditions.

In  finite-sized systems, one can use DF to approximate the J/K matrix and the MO
integrals for the HF, DFT and MP2 methods.  To improve the performance of
the CIAH algorithm, one can use the DF orbital Hessian in the CIAH
orbital optimization for Edmiston-Ruedenberg localization and for the HF, DFT and
CASSCF algorithms.

In the PBC module, the 2-electron integrals are represented as the product of
two 3-index tensors which are treated as DF objects.
Based on the requirements of the system being modelled, we have developed various DF 
representations.
When the calculation involves only smooth bases (typically with pseudopotentials),
plane waves are used as the auxiliary fitting functions and the DF 3-index
tensor is computed within a grid-based treatment using discrete fast Fourier transforms~\cite{McClain2017}.
When high accuracy in all-electron calculations is required,
a mixed density fitting technique is invoked in which the fitting functions are
Gaussian functions plus plane waves.
Besides the choice of fitting basis, different metrics (e.g.~overlap, kinetic, or Coulomb) can be used
in the fitting to balance performance against computational accuracy.

The 3-index DF tensor is stored as a giant array in the HDF5 format without compression.
With this design, it is straightforward to access the 2-electron integrals
with the functions of the \textsc{PySCF} package.
Moreover, it allows us to supply 2-electron integrals to calculations by
overloading the DF object in cases where direct storage of the 4-index integrals in memory or on disk is infeasible
(see discussion in Section \ref{sec:customH}).

\subsection{\sffamily \normalsize Custom Hamiltonians}
\label{sec:customH}
Most quantum chemistry approximations are not tied to the details of the ab initio molecular or periodic Hamiltonian.
This means that they can also be used with arbitrary model Hamiltonians,
which is of interest for semi-empirical quantum chemistry calculations
as well as condensed-matter  model studies.
In \textsc{PySCF}, overwriting the predefined Hamiltonian is straightforward.
The Hamiltonian is an attribute of the mean-field calculation object.
Once the 1-particle and 2-particle integral attributes of the mean-field object are
defined, they are used by the mean-field calculation and all subsequent
post-Hartree-Fock correlation treatments.
Users can thus carry out correlated calculations with model Hamiltonians in exactly the same way as
with standard ab initio Hamiltonians.
Figure~\ref{fig:inputH} displays an example of how to input a model Hamiltonian.

\subsection{\sffamily \normalsize Interfaces to external programs}
\textsc{PySCF} can be used either as the driver to execute external programs or as
an independent solver to use as part of a computational workflow involving other software.
In \textsc{PySCF}, the DMRG programs \textsc{Block}\cite{Sharma2012} and
\textsc{CheMPS2}\cite{CheMPS2cite1,CheMPS2cite4} and the FCIQMC program \textsc{NECI}\cite{neci} can be
used as a replacement for the FCI routine for large active spaces in the
CASCI/CASSCF solver.
In the QM/MM interface, by supplying the charges and the positions of the MM
atoms, one can compute the HF, DFT, MP2, CC, CI and MCSCF energies and their
analytic nuclear gradients.

To communicate with other quantum chemistry programs,
we provide utility functions to read and write Hamiltonians in the
\textsc{Molpro}\cite{MOLPRO} \verb$FCIDUMP$ format, and arbitrary orbitals in the
\verb$molden$\cite{molden} format.
The program also supports to write SCF solution and CI wavefunction in the
\textsc{GAMESS}\cite{gamess} \verb$WFN$ format and to read orbitals from
Molpro XML output.
The real space electron density can be output on cubic grids in the
\textsc{Gaussian}\cite{gaussian} \verb$cube$ format.

\subsection{\sffamily \normalsize Numerical tools}
\label{sec:tools}
Although the \textsc{Numpy} and \textsc{Scipy} libraries provide a wide range of
numerical tools for scientific computing, there are some numerical components
commonly found in quantum chemistry algorithms that are not provided by these
libraries.  For example, the direct inversion of the iterative
space (DIIS) method\cite{Pulay1980,Pulay1982} is one of the most commonly used 
tools in quantum chemistry
to speed up optimizations when a second order algorithm is not
available.
In  \textsc{PySCF} we provide a general DIIS handler for an object array of
arbitrary size and arbitrary data type.  In the current implementation,
it supports  DIIS optimization both with or without supplying the error vectors.
For the latter case, the differences between the arrays of adjacent iterations are minimized.
Large scale eigenvalue and linear equation solvers are also
common components of many quantum chemistry methods.
The Davidson diagonalization algorithm and Arnoldi/Krylov subspace solver
are accessible in \textsc{PySCF} through simple APIs.

\section{\sffamily \Large Design and implementation of \textsc{PySCF}}

While we have tried to provide rich functionality for quantum chemical simulations
with the built-in functions of the \textsc{PySCF} package, it will
nonetheless often be the case that a user's needs are not covered by the built-in functionality.
A major design goal has been to implement \textsc{PySCF} in a sufficiently flexible way so
that users can easily extend its functionality.  
To provide robust components for complex problems and non-trivial workflows,
we have made the following general design choices in
\textsc{PySCF}:
\begin{enumerate}
  \item {\it Language}: Mostly Python, with a little C.  We believe that it is easiest to
  develop and test new functionality in Python.  For this reason, most functions in
  \textsc{PySCF} are written in pure Python.  Only a few computational hot spots have been
  rewritten and optimized in C.
  \item {\it Style}: Mostly functional, with a little object-oriented programming (OOP).
  Although OOP is a successful and widely used programming paradigm, we feel that it is
  hard for users to customize typical OOP programs without learning details of the
  object hierarchy and interfaces.  We have adopted a functional programming style, where
  most functions are pure, and thus can be invoked alone and independently of each other.
  This allows users to mix functionality with a minimal knowledge of the
  \textsc{PySCF} internals.
\end{enumerate}
We elaborate on these choices below.

\subsection{\sffamily \normalsize Input language}

\begin{figure}[htp]
\centering
\begin{minipage}[b]{.5\textwidth}
\begin{pyscf}
import numpy as np
from pyscf import gto, scf
bond = np.arange(0.8, 5.0, .1)
dm_init = None
e_hf = []
for r in reversed(bond):
  mol = gto.Mole(atom=[['N', 0, 0, 0],
                       ['N', 0, 0, r]],
                 basis='ccpvtz')
  mf = scf.RHF(mol).run(dm_init)
  dm_init = mf.make_rdm1()
  e_hf.append(mf.e_tot)

from matplotlib import pyplot
pyplot.plot(bond, e_hf[::-1])
pyplot.show()
\end{pyscf}
\end{minipage}
\begin{minipage}[b]{.48\textwidth}
  \hspace{-6ex}
  \includegraphics[width=1.\textwidth]{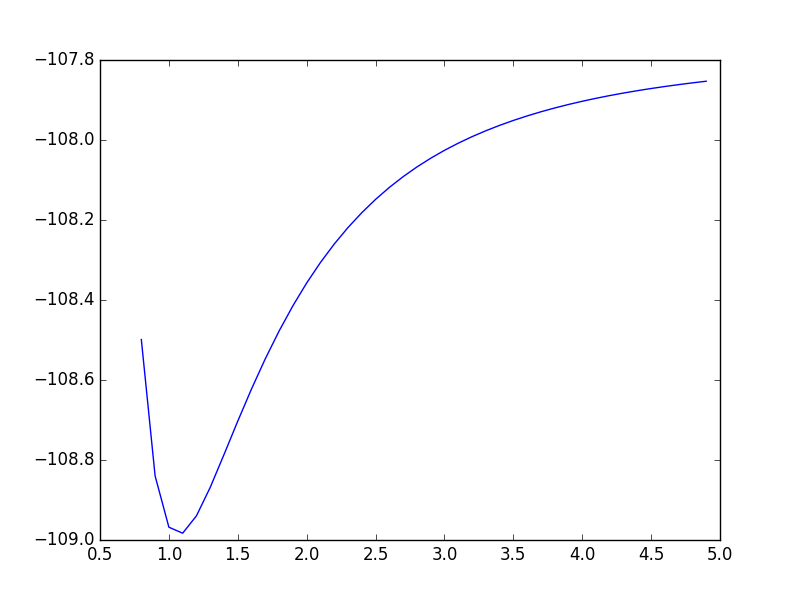}
\end{minipage}
  \caption{Using Python to combine the calculation and data post-processing in one script.}
  \label{fig:workflow}
\end{figure}

Almost every  quantum chemistry package today uses its own custom input language.
This is a burden to the user, who must become
familiar with a new domain-specific language for every new package.
In contrast, \textsc{PySCF} does not have an input language.  Rather, the functionality
is simply called from an input script written in the host Python language. 
This choice has clear benefits:
\begin{enumerate}
  \item {\it There is no need to learn a domain-specific language.}
    Python, as a general programming language, is already widely used for
    numerical computing, and is taught in modern computer science courses.
    For novices, the language is easy to learn and help is readily available from the
    large Python community.
  \item {\it One can use all Python language features in the input script.}
    This allows the input script to implement complex logic and computational
    workflows, and to carry out tasks (e.g.~data processing and plotting) in the same
    script as the electronic structure simulation (see Figure~\ref{fig:workflow} for an
    example).
  \item {\it The computational environment is easily extended beyond that provided by the \textsc{PySCF} package.}
    The \textsc{PySCF} package is a regular Python module
    which can be mixed and matched with other Python modules to build a personalized
     computing environment.
  \item {\it Computing can be carried out interactively.}  Simulations can be tested, debugged, and executed step by step within
    the Python interpreter shell.
\end{enumerate}

\subsection{\sffamily \normalsize Enabling interactive computing}

\begin{figure}[htp]
\centering
  \includegraphics[width=1.\textwidth]{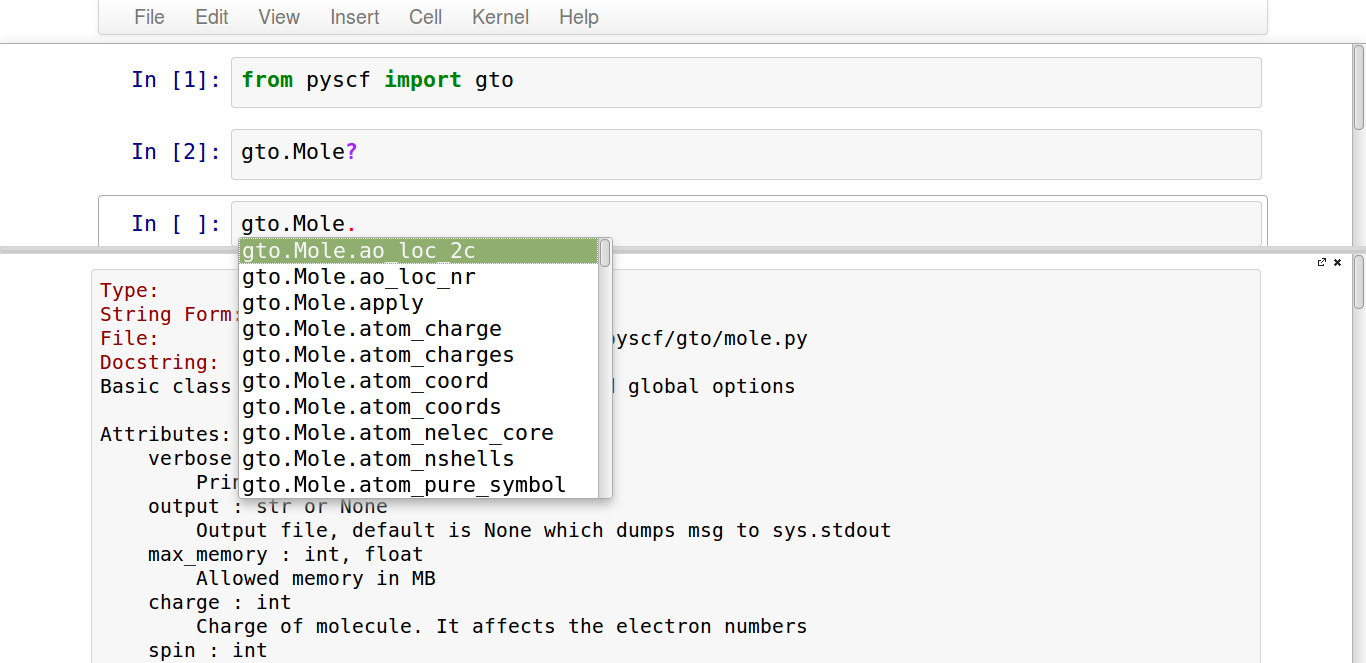}
  \cprotect\caption{Accessing documentation within the IPython shell.
  The question mark activates the documentation window in the bottom area.
  The pop-up menu for code auto-completion is triggered by the \verb$<Tab>$ key.}
  \label{fig:repldoc}
\end{figure}

As discussed above, a strength of the \textsc{PySCF} package is that its functionality
can be invoked from the interactive Python shell. However, maximizing its usability
in this interactive mode entails additional design optimizations.
There are three critical considerations to facilitate such interactive computations:
\begin{enumerate}
  \item The functions and data need to be easy to access;
  \item Functions should be insensitive to execution order (when and how many times
    a function is called should not affect the result);
  \item Computations should not cause (significant) halts in the interactive shell.
\end{enumerate}
To address these requirements, we have enforced the following design rules 
wherever possible in the package:
\begin{enumerate}
  \item Functions are pure (i.e.~state free). This ensures that they are insensitive to execution order;
  \item Method objects (classes) only hold results and control parameters;
  \item There is no initialization of functions, or at most a short initialization chain;
  \item Methods are placed at both the module level and within classes so that
    the methods and their documentation can be easily accessed by the
    interactive shell (see Figure~\ref{fig:repldoc}).
\end{enumerate}
A practical solution to eliminate halting of the interactive shell is to overlap the REPL
(read-eval-print-loop) and task execution.  Such task parallelism requires
the underlying tasks to be independent of each other.  Although certain dependence between methods
is inevitable, the above design rules greatly reduce function call dependence.  Most functions in
\textsc{PySCF} can be safely placed in the background using the standard Python  \verb$threading$ and \verb$multiprocessing$ libraries.

\subsection{\sffamily \normalsize Methods as plugins}
\label{sec:plugin}
Ease-of-use is the primary design objective of the \textsc{PySCF} package.
However, function simplicity and versatility are difficult to balance in
the same software framework.
To balance readability and complexity,
we have implemented only the basic algorithmic features in the main methods, and placed advanced
features in additional ``plugins''.
For instance, the main mean-field module implements only the basic self-consistent loop.
Corrections (such as for relativistic effects) are implemented in an independent
plugin module, which can be activated by reassigning the mean-field 1-electron Hamiltonian
method at runtime.
Although this design increases the complexity of implementation of the plugin functions,
the core methods retain a clear structure  and are easy to comprehend.
Further, this approach decreases the coupling between different features: for example,
independent features can be modified and tested independently and combined in calculations.
In the package, this plugin design has been widely used, for example, to enable molecular point group symmetry,
relativistic corrections, solvation effects, density fitting approximations,
the use of second-order orbital optimization, different variational active space solvers, and many
other features (Figure~\ref{fig:plugin}).

\subsection{\sffamily \normalsize Seamless MPI functionality}
The Message Passing Interface (MPI) is the most popular parallel protocol in the
field of high performance computing.  Although MPI provides high efficiency
for parallel programming, it is a challenge to develop a simple and efficient MPI
program.  In compiled languages, the program must explicitly control 
data communication according to the MPI communication protocol.
The most common design is to activate MPI communication from the beginning and
to update the status of the MPI communicator throughout the program.
When developing new methods, this often leads to extra effort in code
development and debugging.
To sustain the simplicity of the \textsc{PySCF} package, we have designed a
different mechanism to execute parallel code with MPI.
We use MPI to start the Python interpreter as a daemon to receive both the
functions and data on the remote nodes.
When a parallel session is activated, the master process sends to the remote Python
daemons both the functions and the data.  The function is decoded remotely and then
executed.
This design allows one to develop code mainly in serial mode and to switch
to the MPI mode only when high performance is required.  Figure \ref{fig:mpi} shows an
example to perform a periodic calculation with and without a parallel session.
Comparing to the serial mode invocation, we see that the user only has to change the
density fitting object to acquire parallel functionality.

\section{\sffamily \Large CONCLUSIONS}

Python and its large collection of third party libraries are helping to revolutionize how
we carry out and implement numerical simulations.  It is potentially much more productive
to solve computational problems within the Python ecosystem because it frees researchers
to work at the highest level of abstraction without worrying about the details of complex
software implementation.  To bring all the benefits of the Python ecosystem to quantum
chemistry and  electronic structure simulations, we have
started the open-source \textsc{PySCF} project.

\textsc{PySCF} is a simple, lightweight, and efficient computational chemistry program
package, which supports ab initio calculations for both molecular and extended systems.
The package serves as an extensible electronic structure toolbox, providing a large number
of fundamental operations with simple APIs to manipulate methods, integrals, and wave
functions.  We have invested significant effort to ensure simplicity of use and
implementation while preserving competitive functionality and performance. We believe that
this package represents a new style of program and library design that will be
representative of future software developments in the field.

\section{\sffamily \Large ACKNOWLEDGMENTS}

QS would like to thank Junbo Lu and Alexander Sokolov for testing functionality and for
useful suggestions for the program package.  The development of different components of
the  \textsc{PySCF} package has been generously supported by several sources.  Most of the
molecular quantum chemistry software infrastructure was developed with support from the US
National Science Foundation, through grants CHE-1650436 and ACI-1657286. The
periodic mean-field infrastructure was developed with support from ACI-1657286. The
excited-state periodic coupled cluster methods were developed with support from the US
Department of Energy, Office of Science, through the grants DE-SC0010530 and DE-SC0008624.
Additional support for the extended-system methods has been provided by the Simons
Foundation through the Simons Collaboration on the Many Electron Problem, a Simons
Investigatorship in Theoretical Physics, the Princeton Center for Theoretical Science, and
startup funds from Princeton University and the California Institute of Technology.



\bibliography{ref}

%
%

\clearpage

\begin{table}
  \centering
  \caption{Features of the PySCF package as of the 1.3 release.}
  \begin{threeparttable}
  \begin{tabular}{llllllllllllll}
    \toprule
    Method    & Molecule & Solids & Comments \\
    \midrule
    HF        & Yes & Yes          & $\sim$ 5000 AOs\tnote{b} \\
    MP2       & Yes & Yes          & $\sim$ 1500 MOs\tnote{b} \\
    CCSD      & Yes & Yes          & $\sim$ 1500 MOs\tnote{b} \\
    EOM-CCSD  & Yes & Yes          & $\sim$ 1500 MOs\tnote{b} \\
    CCSD(T)   & Yes & Yes\tnote{a} & $\sim$ 1500 MOs\tnote{b} \\
    MCSCF     & Yes & Yes\tnote{a} & $\sim$ 3000 AOs\tnote{b}, \ 30--50 active orbitals\tnote{c} \\
    MRPT      & Yes & Yes\tnote{a} & $\sim$ 1500 MOs\tnote{b}, \ 30--50 active orbitals\tnote{c} \\
    DFT       & Yes & Yes          & $\sim$ 5000 AOs\tnote{b} \\
    TDDFT     & Yes & Yes\tnote{a} & $\sim$ 5000 AOs\tnote{b} \\
    CISD      & Yes & Yes\tnote{a} & $\sim$ 1500 MOs\tnote{b} \\
    FCI       & Yes & Yes\tnote{a} & $\sim$ (18e, 18o)\tnote{b} \\
    \multirow{2}{*}{Localizer} & \multirow{2}{*}{Yes} & \multirow{2}{*}{No}
                                   & IAO, NAO, meta-L\"owdin \\
              &     &              & Boys, Edmiston-Ruedenberg, Pipek-Mezey\\
    \multirow{2}{*}{Relativity}& \multirow{2}{*}{Yes} & \multirow{2}{*}{No}           & ECP and scalar-relativistic corrections for all methods \\
              &     &              & 2-component, 4-component methods for HF and MP2 \\
    Gradients & Yes & No           & HF, DFT, CCSD, CCSD(T), TDDFT \\
    Hessian   & Yes & No           & HF and DFT \\
    Properties  & Yes & No           & non-relativistic, 4-component relativistic NMR \\
    Symmetry  & Yes & No           & D$_{2h}$ and subgroup \\
    AO, MO integrals & Yes & Yes   & 1-electron, 2-electron integrals \\
    Density fitting & Yes & Yes    & HF, DFT, MP2 \\
    \bottomrule
  \end{tabular}
\begin{tablenotes}
\item[a] Only available for $\Gamma$-point calculations;
\item[b] An estimation based on a single SMP node with 128 GB memory without density fitting;
\item[c] Using an external DMRG or FCIQMC program as active space solver.
\end{tablenotes}
  \end{threeparttable}
  \label{tab:features}
\end{table}

\clearpage

\begin{figure}[htp]
\centering
\begin{pyscf}
from pyscf import gto, dft
mol = gto.Mole(atom='N 0 0 0; N 0 0 1.1', basis='ccpvtz')
mf = dft.RKS(mol)
mf.xc = '0.2*HF + 0.08*LDA + 0.72*B88, 0.81*LYP + 0.19*VWN'
mf.kernel()
\end{pyscf}
  \caption{Example to define a custom exchange-correlation functional for a DFT calculation.}
  \label{fig:xc}
\end{figure}

\begin{figure}[htp]
\centering
\begin{pyscf}
from pyscf import gto, scf, mcscf, dmrgscf
mol = gto.Mole(atom='N 0 0 0; N 0 0 1.1', basis='ccpvtz')
mf = scf.RHF(mol).run()
mc = mcscf.CASSCF(mf, 8, 10)  # 8o, 10e
mc.fcisolver = dmrgscf.DMRGCI(mol)
mc.kernel()
\end{pyscf}
  \caption{Example to enable the DMRG solver in a CASSCF calculation.}
  \label{fig:dmrgscf}
\end{figure}

\begin{figure}[htp]
\centering
\begin{pyscf}
from pyscf import gto
mol = gto.Mole(atom='N 0 0 0; N 0 0 1.1', basis='ccpvtz')
a = mol.intor('cint1e_nuc_sph')  # nuclear attraction as a giant array
a = mol.intor('cint2e_sph')      # 2e integrals as a giant array
a = mol.intor_by_shell('cint2e_sph', (0,0,0,0))  # (ss|ss) of first N atom
\end{pyscf}
  \caption{Example to access AO integrals.}
  \label{fig:ao:ints}
\end{figure}

\begin{figure}[h]
\centering
\begin{pyscf}
# 10-site Hubbard model at half-filling with U/t = 4
import numpy as np
from pyscf import gto, scf, ao2mo, cc
mol = gto.Mole(verbose=4)
mol.nelectron = n = 10
t, u = 1., 4.

mf = scf.RHF(mol)
h1 = np.zeros((n,n))
for i in range(n-1):
    h1[i,i+1] = h1[i+1,i] = t
mf.get_hcore = lambda *args: h1
mf.get_ovlp = lambda *args: np.eye(n)
mf._eri = np.zeros((n,n,n,n))
for i in range(n):
    mf._eri[i,i,i,i] = u
# 2e Hamiltonian in 4-fold symmetry
mf._eri = ao2mo.restore(4, mf._eri, n)
mf.run()
cc.CCSD(mf).run()
\end{pyscf}
  \caption{Example to use a custom Hamiltonian.}
  \label{fig:inputH}
\end{figure}

\begin{figure}[htp]
\centering
\begin{pyscf}
from pyscf import gto, scf
mol = gto.Mole(atom='N 0 0 0; N 0 0 1.1', basis='ccpvtz')
mf = scf.newton(scf.sfx2c(scf.density_fit(scf.RHF(mol)))).run()
\end{pyscf}
  \caption{Example to use plugins in PySCF.  The mean-field calculation is
  decorated by the density fitting approximation, X2C relativistic correction
  and second order SCF solver.}
  \label{fig:plugin}
\end{figure}

\begin{figure}[htp]
\centering
\begin{minipage}[c]{.41\textwidth}
\centering
\begin{pyscf}
# Serial mode
# run in cmdline:  
# python input.py
from pyscf.pbc import gto, scf
**from pyscf.pbc import df**
cell = gto.Cell()
cell.atom = 'H 0 0 0; H 0 0 0.7'
cell.basis = 'ccpvdz'
# unit cell lattice vectors
cell.a = '2 0 0; 0 2 0; 0 0 2'
# grid for numerical integration
cell.gs = [10,10,10]
mf = scf.RHF(cell)
mf.with_df = df.DF(cell)
mf.kernel()
\end{pyscf}
\end{minipage}
\begin{minipage}[c]{.45\textwidth}
\centering
\begin{pyscf}
# MPI mode
# run in cmdline:  
# mpirun -np 4 python input.py
from pyscf.pbc import gto, scf
**from mpi4pyscf.pbc import df**
cell = gto.Cell()
cell.atom = 'H 0 0 0; H 0 0 0.7'
cell.basis = 'ccpvdz'
# unit cell lattice vectors
cell.a = '2 0 0; 0 2 0; 0 0 2'
# grid for numerical integration
cell.gs = [10,10,10]
mf = scf.RHF(cell)
mf.with_df = df.DF(cell)
mf.kernel()
\end{pyscf}
\end{minipage}
  \caption{Comparison of the input script for serial-mode and MPI-mode
  calculations.  Except for the module to import, the MPI parallel mode takes exactly
  the same input as the serial mode.}
  \label{fig:mpi}
\end{figure}

\end{document}